%% file: main.tex
\documentclass[sigconf]{acmart}

\graphicspath{{./images/}} 

\copyrightyear{2024}
\acmYear{2024}
\setcopyright{rightsretained}
\acmConference[CSCW Companion '24]{Companion of the 2024
Computer-Supported Cooperative Work and Social Computing}{November 9--13,
2024}{San Jose, Costa Rica}
\acmBooktitle{Companion of the 2024 Computer-Supported Cooperative Work
and Social Computing (CSCW Companion '24), November 9--13, 2024, San Jose,
Costa Rica}\acmDOI{10.1145/3678884.3681866}
\acmISBN{979-8-4007-1114-5/24/11}

\begin{document}

\title[Assembling the Puzzle]{Assembling the Puzzle: Exploring Collaboration and Data Sensemaking in Nursing Practices for Remote Patient Monitoring}


\author{Mihnea Stefan Calota}
\email{m.s.calota1@tue.nl}
\orcid{0009-0002-5979-2439}
\affiliation{%
  \institution{Eindhoven University of Technology}
  \city{Eindhoven}
  \country{The Netherlands}
}

\author{Janet Yi-Ching Huang}
\email{y.c.huang@tue.nl}
\orcid{0000-0002-8204-4327}
\affiliation{%
  \institution{Eindhoven University of Technology}
  \city{Eindhoven}
  \country{The Netherlands}
}

\author{Lin-Lin Chen}
\email{l.chen@tue.nl}
\orcid{0000-0001-9887-9858}
\affiliation{%
  \institution{Eindhoven University of Technology}
  \city{Eindhoven}
  \country{The Netherlands}
}

\author{Mathias Funk}
\email{m.funk@tue.nl}
\orcid{0000-0001-5877-2802}
\affiliation{%
  \institution{Eindhoven University of Technology}
  \city{Eindhoven}
  \country{The Netherlands}
}

\renewcommand{\shortauthors}{Mihnea Stefan Calota, Janet Yi-Ching Huang, Lin-Lin Chen, Mathias Funk}

\begin{abstract}
\input{main_poster/0_Abstract}
\end{abstract}



\begin{teaserfigure}
  \includegraphics[width=\textwidth]{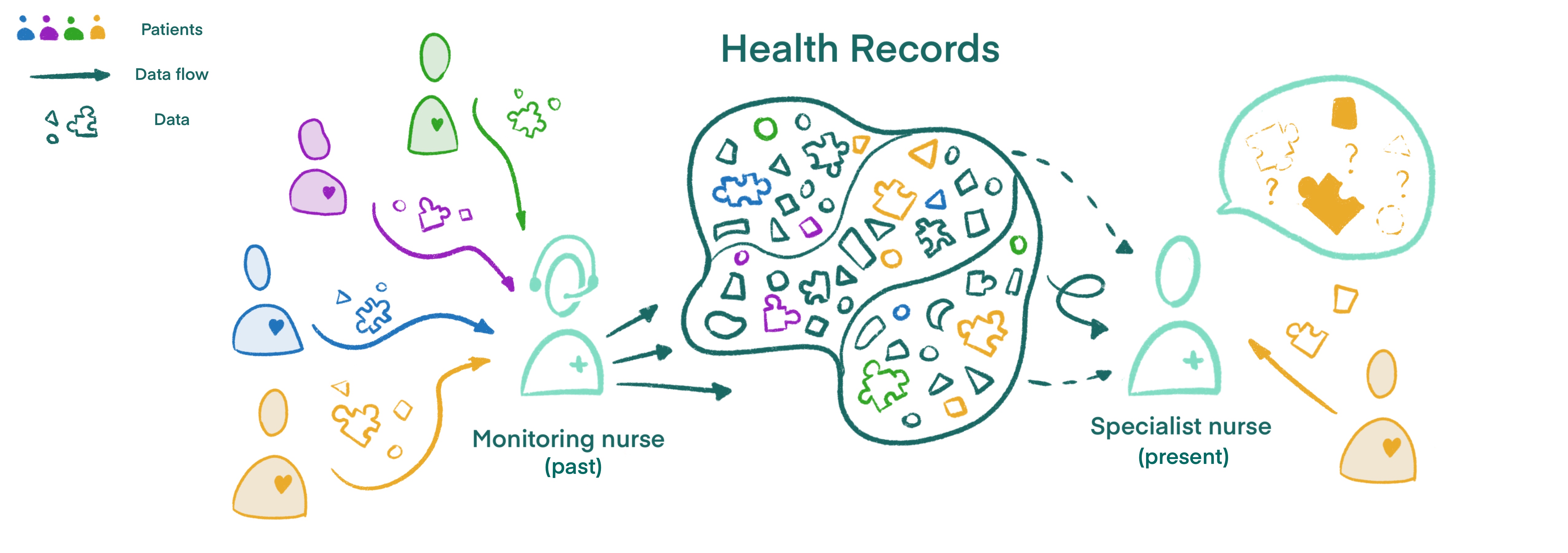}
  \caption{The knowledge transfer between a past monitoring nurse (left) and a present specialist nurse (right) is interfaced through the health records system, leading to asynchronous collaboration. The monitoring nurse collected patient data (green, purple, blue, yellow) and stored it in the health records system.  The specialist nurse needs information about a specific patient (yellow), and collects some data from the patient directly. They still need to extract data from the health records system. Finding and decoding this data forms a ``data sensemaking'' activity. This task creates an increased cognitive load and friction in the workflow of the nurse.}
  \Description{}
  \label{fig:RPM collab}
\end{teaserfigure}

\maketitle

\input{main_poster/1_Introduction}

\input{main_poster/2_Methods}
\input{main_poster/3_Results}
\input{main_poster/4_Discussion}
\input{main_poster/5_Conclusion}


\section*{Acknowledgements}
We would like to thank all of our participants for allowing us a glimpse into their day-to-day work and for taking time from their busy schedules to answer our questions.

\bibliographystyle{ACM-Reference-Format}
\balance
\bibliography{zotero.bib}

\end{document}

%% file: main_poster/0_Abstract.tex
Remote patient monitoring (RPM) involves the remote collection and transmission of patient health data, serving as a notable application of data-driven healthcare. This technology facilitates clinical monitoring and decision-making, offering benefits like reduced healthcare costs and improved patient outcomes. However, RPM also introduces challenges common to data-driven healthcare, such as additional data work that can disrupt clinician's workflow. This study explores the daily practices, collaboration mechanisms, and sensemaking processes of nurses in RPM through field observations and interviews with six stakeholders. Preliminary results indicate that RPM's scale-up pushes clinicians toward asynchronous collaboration. Data sensemaking is crucial for this type of collaboration, but existing technologies often create friction rather than support. This work provides empirical insights into clinical workflow in nursing practice, especially RPM. We suggest recognizing data sensemaking as a distinct nursing practice within data work and recommend further investigation into its role in the workflow of nurses in RPM.

%% file: main_poster/1_Introduction.tex
\section{Introduction}  

The healthcare sector is constantly changing, with new technologies and methodologies implemented daily. Data-driven healthcare refers to the practice of using tools and technology to collect, analyze, and leverage patient data to improve healthcare outcomes, enhance decision-making, and optimize clinical workflows. Leveraging patient data allows clinicians to tailor the care programs to each individual's needs or to use predictive analysis to intervene early and reduce hospital (re)admissions \cite{grossglauser_data-driven_2014}. In turn, this can lead to lower costs for healthcare providers and improved quality of care for patients~\cite{amri_data-driven_2023}. 

However, data-driven healthcare technologies often face tremendous challenges in transitioning from the design stage to real-world implementation \cite{yang_unremarkable_2019}. These technologies often fail in practice due to poor contextual fit or disruption of established clinical workflows \cite{yang_unremarkable_2019,serban_i_2023}. For instance, a tool that does not integrate seamlessly with existing electronic systems can create additional administrative burdens for clinicians, distracting them from patient care. Similarly, technologies that do not align with the routines and needs of healthcare professionals are often underutilized or misused, negating their potential benefits. To ensure successful implementation and adoption, it is important to gain a deeper understanding of the specific contexts within which these technologies will be used.

One notable application of data-driven healthcare is Remote Patient Monitoring (RPM). It involves the intensive use of technology to collect and transmit patient data (i.e., physiological measurements and symptoms) from a remote location (the patient's home) to the clinicians for interpretation and decision-making. In turn, clinicians provide remote treatment through advice, coaching, or medication adjustments based on the insights derived from the provided data and their professional experience. RPM was shown to offer numerous advantages, including improved quality of life for patients, reduced healthcare costs, and a decrease in hospitalization needs \cite{olen_implantable_2017}. However, the remote nature of RPM requires a shift in the way clinicians work, involving considerable data work \cite{sun_data_2023} and distributing collaboration between clinicians over space and time. 

Data-driven healthcare, and RPM as a subset of it, is inherently connected to data work conducted most often by clinicians. They handle data gathering, storage, retrieval, and sensemaking \cite{bertelsen_data_2024}. Understanding how clinicians react to the changes in their workflow imposed by data work is essential to the successful implementation of RPM programs. Previous studies have investigated various aspects of designing clinician support tools \cite{yang_unremarkable_2019,sultanum_more_2018} and their impact on the clinician workflow \cite{devaraj_barriers_2014}. However, most focus on operational constraints or tangible measures of workflow efficiency, with little work trying to unpack data work and cognitive activities clinicians carry out. Our research aims to investigate data work activities in RPM to further understand how they influence clinicians' workflow and collaboration strategies.

\subsection{The interplay of human and technological factors in RPM} 

Specially trained nurses play an essential role in keeping an RPM program running smoothly. They act as an interface between patients, data records, and other clinicians. However, their key placement in the RPM system leads to nurses experiencing some of the most disruptive changes to their pre-RPM workflows, as their jobs are highly collaborative and interconnected with the management and interpretation of patient data \cite{grisot_data-work_2019}. Despite the importance of nurse-led monitoring programs, prior studies often focus more on the patient rather than the nurse experience \cite{auton_smartphone-based_2023}. 

The implementation of new technologies in RPM requires nurses to adapt many facets of their usual clinical practice: face-to-face consults become phone calls, collaboration with colleagues happens remotely or asynchronously, measurements are taken by the patients themselves, and new data types coming from new sources need be filtered, cleaned and filed in the same unchanged electronic health record (EHR)\cite{maslen_layers_2017}\cite{varpio_ehr_2015}\cite{sun_data_2023}. In this context, Grisot et al. emphasize the critical role specially trained nurses (i.e., monitoring nurses) play in remote monitoring. Their study analyzes novel nursing practices that emerge from RPM implementation \cite{grisot_data-work_2019}. Their results show that data work conducted by specialized monitoring nurses is key to maintaining high-quality remote healthcare.

Therefore, we consider the monitoring nurse to be one of the central stakeholders in the RPM system. We aim to look at the RPM workflow through the nurse's lens to understand their experience, particularly their routine practice, perception, and challenges surrounding data work and data-driven healthcare.  

\subsection{Data work for RPM} 
Data work is a concept used to understand nurse practice in RPM. This study defines data work as any human activity related to creating, collecting, managing, curating, analyzing, interpreting, and communicating data \cite{sun_data_2023}. Data is not simply ready to be harvested but must be created, managed, and formatted through intentional effort~\cite{bossen_data_2019}. This is especially relevant in healthcare, where the desire to benefit from ``data-driven'' healthcare often clashes with the reality that intense data work will be carried out, most often by clinicians.  

In the context of RPM, Islind et al. report on how nurses in the monitoring center translate and make data useful \cite{islind_shift_2019}. Their study focuses on three shifts in nursing practice stemming from the introduction of patient-generated data as a core decision-making material: a shift in questioning tactics, a shift in work distribution, and, most importantly, a shift in decision-making.

The shift in questioning tactics stems from the physical distance between the clinician and the patient, imposed by the remote nature of RPM. Successful diagnosis depends upon the clinician's ability to perceive changes in the patient, and therefore, they have to ask them about things they cannot see in the data \cite{maslen_layers_2017}.

The shift in work distribution aligns with the RPM workflow, where continuous data work is incorporated into the daily routine of a nurse. This contrasts with the traditional model, which primarily involved one-time data effort during face-to-face consultations. 


The shifts in work distribution and questioning strategy, with the addition of data work, also directly influence the decision-making processes of clinicians \cite{islind_shift_2019}. This is because nurses rely on browsing and making sense of the data to make their decisions. While technological solutions like EHRs and monitoring apps (such as Luscii~\cite{noauthor_luscii_nodate}) are being developed to support these shifts towards personalized healthcare, their actual impact on nurse practice and the level of support they provide remain unclear~\cite{varpio_ehr_2015}\cite{wright_you_2015}.

In personalized healthcare, three key nurse practices related to data work have been identified: preparation activities, continuous adjustments, and questionnaire fine-tuning~\cite{grisot_data-work_2019}. Sensemaking activities, which are critical yet dispersed across all three practices, are particularly notable. We propose positioning data sensemkaing as a distinct nurse practice within data work and exploring its role in the workflow of nurses in RPM.  

\begin{table*}[t]
    \centering
    \caption{Participant Data}\label{PD}
    \begin{tabular}{|c|c|c|c|c|}
    \hline
    \textbf{Patient ID} & \textbf{Role} & \textbf{Institution} & \textbf{Type of Study}& \textbf{Duration} \\
    \hline
    P1 & Specialist Nurse & Hospital 1 & Observation \& Interview & 4 hours (including interview) \\
    P2 & Monitoring Nurse & Hospital 1 & 2 Observations \& Interview & 5 hours (including interview) \& 4 hours \\
    P3 & Project Lead & Hospital 1 & Interview & 1 hour \\
    P4 & Clinician \& Researcher & Hospital 1 & Observation \& Interview  & 2 hours (including interview)\\
    P5 & Monitoring Nurse & Hospital 2 & Interview & 1 hour \\
    P6 & Monitoring Nurse & Hospital 2 & Interview & 1.5 hours \\
    \hline
    \end{tabular}
\end{table*}

\subsection{(Data) sensemaking} 
Sensemaking research has been explored across a variety of fields and participant types, but it constantly focuses on how humans transform ambiguous and complex situations into understandable and actionable information \cite{kaplan_framing_2008, sandberg_making_2015, maitlis_sensemaking_2014}. In areas like user interface design, sensemaking is regarded as an independent framework used to analyze the cognitive processes of users \cite{suh_sensecape_2023,fisher_distributed_2012}. In RPM, a nurse is one part of a complex system, transforming intricate and disconnected data points into decisions and actions. In this context, a nurse is embedded in a data-driven environment, and data work and sensemaking become integral to their professional identity \cite{korica_making_2010}.

In our study, sensemaking activities are seen as cognitive and sometimes collaborative processes where varied data sources (i.e., patient measurements, EHR data, phone calls) are interpreted to inform personalized care and clinical decision-making. Data sensemaking is not entirely an internal process but is influenced by material props, in this case, data \cite{stigliani_organizing_2012}. This perspective is based on the work of Suh et al. on ``Sensecape'' \cite{suh_sensecape_2023}, and the work of Hultin \cite{hultin_how_2017} on how sensemaking occurs in the emergency room. The primary goal of data sensemaking for clinicians is to transform discrete information pieces into coherent, actionable insights.  

Despite the advantages of RPM in improving patient outcomes and reducing healthcare costs, significant challenges remain in integrating data work into clinical workflows. Previous research often overlooked the cognitive and collaborative activities nurses perform to manage and interpret patient data. This gap highlights the need to explore data work practices and their impact on nurse experience. Addressing this gap is crucial for developing data-driven solutions that support nurses without disrupting workflows. This study aims to fill this void by examining the specific data work processes and collaborative practices of nurses in the RPM settings.

%% file: main_poster/2_Methods.tex
\section{Method}
This study aims to explore and understand how data-driven characteristics of the RPM environment influence nurse practices and collaboration strategies. We explore two specific questions: \textit{how do nurses in an RPM setting engage in collaboration} and \textit{how does the available technology at their disposal offer support or create friction}.

\subsubsection*{Study design and participants}
In this exploratory study, we recruited six participants from two hospitals in the Netherlands. Both hospitals employ a \emph{commercial} state-of-the-art remote monitoring system, relying on clinicians (i.e., monitoring nurses, specialist nurses, and physicians) to provide remote care using health records, patient data, and patient-generated measurements recorded in the Luscii mobile app \cite{noauthor_luscii_nodate}. The participants had varied roles in the RPM workflow, as shown in Table \ref{PD}. The two institutions had different scales of RPM operation, with Hospital 1 having around 200 patients in 3 care pathways and Hospital 2 having around 2500 patients in over 15 care pathways.

\subsubsection*{Data Collection}
We used field observations and semi-structured interviews to understand the participants' personal and professional contexts and their experiences in RPM.  

As prior work has argued, immersing researchers in the actual work environment of nurses can yield valuable insights into their daily tasks, interactions, and challenges \cite{roth_using_2000}. Observations were conducted in the mornings, aligning with what the participants regarded as peak activity times for RPM interactions. We initially discussed the expectations for the day with the nurses and then minimized interaction to avoid influencing their behaviors. Detailed notes, including timestamps of various actions, were recorded throughout each observation session. The sessions were conducted independently, aligning with the clinical specialists' shifts.

After the observations, we conducted semi-structured interviews with the six participants. These interviews focused on identifying the types of activities the nurses performed during RPM sessions, their perceptions of the data work, and any challenges they faced.

\subsubsection*{Data Analysis}
We qualitatively analyzed data from field notes and interview notes. We used affinity diagrams to organize and categorize the data, which helped identify common themes and patterns in the nurses' activities and interactions during RPM \cite{lucero_using_2015}. We also created timelines and workflow journey maps from the observational notes, which visually represent the nurses' tasks and decision-making processes across typical RPM sessions. These visualizations allowed us to examine how different activities overlap, how collaboration takes place, and where various data-related tasks are performed throughout an average workday. A streamlined example of such timelines can be seen in Figure \ref{fig:nurse timeline}.

%% file: main_poster/3_Results.tex
\section{Results}
The different interviews and observation sessions offered a chance to understand the work environment surrounding RPM. Three main themes emerged: the role of the monitoring nurse in the RPM system, the asynchronous nature of remote care, and the role data sensemaking plays in collaboration.

\subsection{The role of the monitoring nurse}
The daily RPM workflow was highly structured, with monitoring nurses primarily managing alerts and patient calls before noon. Unresolved cases were then transferred to specialist nurses, who investigated, made decisions, and communicated these back to the monitoring nurse. In cases of particular complexity, a specialized physician was consulted. According to P6, out of approximately 700 alerts at the start of the day, around 15\% of them will be seen by a specialist nurse, and only a few will require the attention of a physician. 

Monitoring nurses performed various tasks, including sorting alerts, contacting patients for clarifications, interpreting data, and resolving technical issues. They handled the majority of the data work and sensemaking activities that kept the RPM system running. 

P2 stated, ''I am doing so much more than just sort through alerts.'' and ''I help patients feel safe and listened to, I solve their technical issues. As a [monitoring] nurse, you need to know a bit of everything.''

\begin{figure*}[thbp]
  \centering
  \includegraphics[width=\textwidth]{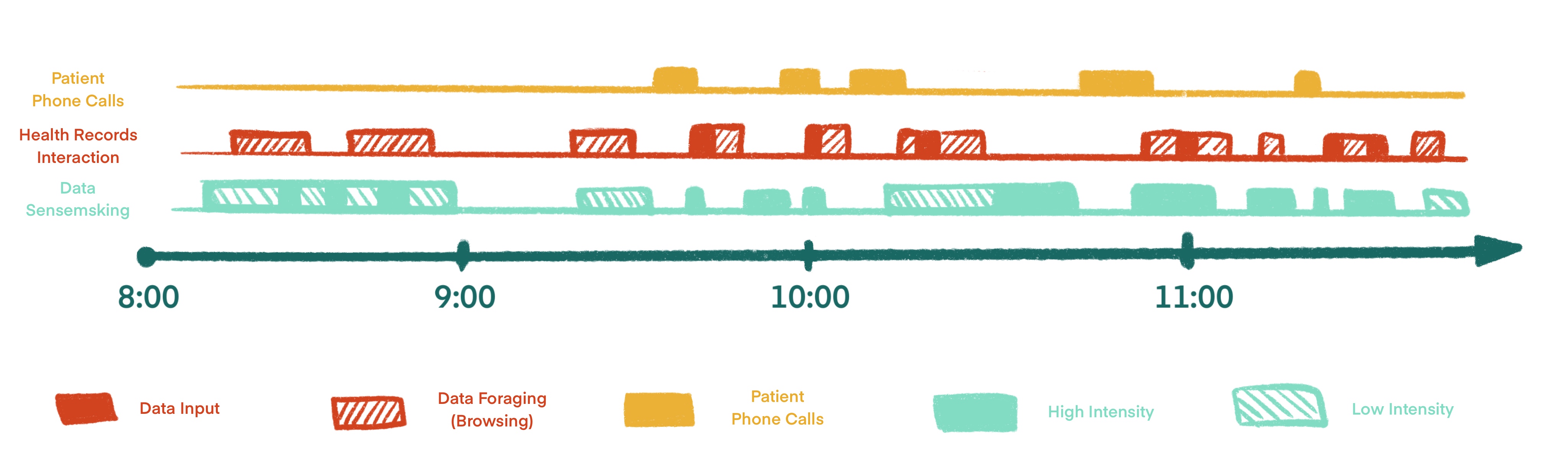}
  \caption{In the daily workflow, a monitoring nurse mainly deals with three types of activities related to sensemaking: calling patients for additionally information (yellow), interacting with the data records (red) and cognitive processes of data sensemaking (green). In a majority of cases, high intensity data sensemaking is correlated with the need to browse or forage for information in the data records.}
  \label{fig:nurse timeline}
\end{figure*}

Despite variations in team sizes and patient loads, the workflow for a single monitoring nurse remained largely consistent. P2 handled all alerts before noon and compiled a report of the patients requiring further discussion with the specialist nurse (P1). As shown in Figure \ref{fig:nurse timeline}, she alternated between calling patients, making sense of different alerts, and browsing the EHR. 

Participants provided diverse responses when asked about the role and value of a monitoring nurse. Some responses included: ''the [monitoring] nurse is there to make the patient feel safe'' (P5), ''to call [the patients], hear their story and understand what they are experiencing'' (P2), ''the [monitoring] nurse is the one that keeps the whole system running'' (P3), ''[their] role is to teach patients to self manage, or at least support the ones who can't do that'' (P6) and ''prevent them from coming back to the hospital.'' (P1)


\subsection{The asynchronous nature of remote care} 
During the observations, we noticed that the remote, data-driven nature of the RPM system shapes collaboration strategies. It encourages frequent and continuous monitoring that requires a relatively low effort in the moment, spread over a prolonged period. As a result, there were almost no traditional intense collaborative efforts. What we often saw were smaller, lower-effort asynchronous collaboration moments. Instead of meetings and live discussions, the nurses typically left notes and small pieces of data in the EHR for later use or communication. 

For instance, P2 noted a patient's state of dizziness in the patient's consult file after a phone call. She decided not to adjust anything and left a note for a colleague to check on the patient later by mentioning ''no medication change for now, to be checked again in two days'' in the file. Later, P1 consulted a patient via a phone call after noticing some inconsistencies in the data submitted through the monitoring app. After the call, P1 still had questions, so he checked the EHR for notes left by his colleagues about this specific patient. This interaction, captured in Figure~\ref{fig:RPM collab}, represents an instance of asynchronous collaboration.

Throughout the observations, we saw both sides of this asynchronous collaboration. As depicted in Figure~\ref{fig:nurse timeline}, the monitoring nurse went back and forth between inputting data for future reference and searching for required data in the system.

Both P4 and P6 mentioned the challenges of this process. ''Most of the time, a written note captures less information than a conversation'' (P6). P4 explained that they need to cross-reference notes with other data points, such as medication, admissions, or measurements, to fully understand them. For example, P4 found a note of a patient having trouble breathing in the morning. This information became more valuable after reviewing previous notes and the patient's admission history. By piecing together these facts, P4 deduced that the patient had previously been admitted to the hospital for influenza and was likely still recovering. 

Lastly, our participants highlighted differences in how the two hospitals organize their RPM programs. The differences were related to how much the clinicians rely on protocols and technology for collaboration. For example, in a smaller RPM program, P1 and P2 could quickly and flexibly collaborate through quick calls and easily understood notes. As P2 noted, ''[P1] and I are working very closely together on the Heart Failure program, so I already know when to call him, when he is available. Sometimes, he just calls early because he has some free time, and I just make the time, and we have a quick chat about a few patients.''

By contrast, monitoring nurses in larger programs (P5 and P6) relied primarily on asynchronous collaboration through reports, notes, and protocols due to the program's large scale. As P6 explained, ''We need to have all the reports and urgent patients ready to go when we contact a specialist nurse. On paper, we have one hour to talk to them around 13:00. In reality, they are very busy, so we just need to have everything organized and placed in the system so they can check [the patients] there when they have more time.''

P3 also anticipated that ''when the program scales up, [they] will just have to do everything by the protocol,'' with technology playing a crucial role in mediating clinician interactions. 

\subsection{The role of data sensemaking in collaboration}

We observed that nurses frequently engage in data sensemaking activities to understand a patient's case. Most of the time, when frustration was visibly expressed (e.g., body language, sighs) or explicitly mentioned, they were trying to locate information in the freeform notes. P2 explained that they could ``pin'' certain information (e.g., smoking behavior) to the top of a patient's profile, eliminating the need to ask for or search for this information in the EHR again. P3 emphasized the significant amount of time spent ``looking for the information you need.''

During the interview, P4 discussed how clinicians are motivated to make their notes as complete as possible. ''You want to have everything in there, both from a legal and insurance perspective and to make sure you didn't miss something important. But everything is just text blocks, and so much information is duplicated. And you have to [search and search] for what you need.'' (P4)

P3 argued that having too much data can cause as much friction as having too little, and P2 observed that the EHR ``is overloaded, showing you everything, all the time, while in reality [the nurse] needs different things at different times, with different patients.'' P5 and P6 agreed, describing the EHR more like a data management system than a support tool. They expressed a desire for a customizable interface that would allow them to selectively view relevant information. 

P1 also saw the data work and sensemaking as potential bottlenecks for scaling up the RPM system and believed that more monitoring nurses would be needed to handle the required data work. In contrast, the monitoring nurses felt that data work is not where their time is best spent, but they ''do it because someone has to do it for now'' (P6). 

%% file: main_poster/4_Discussion.tex
\section{Discussion}
RPM represents a socio-technical system that includes patients, nurses, physicians, communication channels, data gathering systems, data storage systems, and the data itself. This study uncovers potential friction points in the RPM system by exploring how nurses engage with data and collaborate with other clinicians.

\subsection{The humans in the RPM workflows}
The monitoring nurses serve as the interface between patients, other clinicians, and the supporting data systems. One finding of this study is that data work emerges in their daily practice and heavily influences their workflows as they deal with higher workloads and new types of tasks they might not have ever performed before. Next to that, data work is spread thinly over almost all activities of the monitoring nurses: they need to ``know a bit of everything,'' and their role extends beyond simply filtering alerts and inputting data into the EHR. Many participants recognized the core value of the monitoring nurse in the interpersonal relations with patients and their ability to guide patients on self-management while providing a sense of safety. While quality data work is essential to keep the RPM system running, nurses prefer not to spend the majority of their effort on this aspect, nor do they feel it has the most impact on the patient's life.

The nurses attend to multiple patients at varying times, but a constant element in their daily practice is their interaction with the EHR, which they use to transmit or retrieve patient information (Figure \ref{fig:nurse timeline}). All participants consistently mentioned the interpretation of patient— or clinician-generated data as a crucial part of their collaboration and job performance. Our results highlight two sources of friction that could affect nurses' experiences in the RPM workflow: the need to shift to asynchronous collaboration and the obstacle of conducting sensemaking activities with currently available technology. 


\subsection{The burden of synchronizing asynchronous data and collaboration in RPM}
Clinicians typically have demanding schedules and often work in parallel with their colleagues, making it difficult to align their schedules. To effectively leverage new technologies like RPM, it is crucial to understand and prioritize clinicians' workflow and context. Previous studies have identified a gap between the design and real-world implementation of data-driven solutions \cite{yang_unremarkable_2019}. These solutions may perform well in design studios, but they can disrupt clinicians' workflows and lack contextual fit in clinical practice. Preliminary findings from this study indicate that while RPM is pushing clinicians to adopt asynchronous collaboration strategies, the design of the available technology does not fully consider this new workflow. 


Asynchronous collaboration has the advantage of freeing clinicians from the constraints of aligning busy schedules. However, this new collaboration style presents challenges, especially when information must be encoded, processed through a central data management system (such as the EHR), and then decoded (Figure \ref{fig:RPM collab}). Our findings highlight that it is important to strike a balance between too little and too much information available at any moment. The participants voiced their dissatisfaction with the current design principle of the EHR, which is having access to everything at all times. Since clinicians tend to take precautions and prefer to be complete in their notes, the EHR is being populated with dense, lengthy notes and information duplicates, which further add difficulty to the sensemaking activities.

This paper offers a unique insight: asynchronous collaboration in the medical field differs from other sectors. For example, when programmers or researchers work on a shared artifact (such as software or a paper), their cooperation focuses on the changes and additions made to the artifact and is outcome-oriented (i.e., finishing the paper)~\cite{altmann_cooperative_1999}. In contrast, clinicians' asynchronous collaboration is focused on a \emph{transfer of knowledge}, spanning various informational systems, patients, or cases. There is no central artifact to track ``progress'' or ``completion'' of the task. While clinicians usually look at the patient's well-being as a measure of success, this metric does not mark an end to the collaboration efforts.


Lastly, the scale of operation is directly proportional to the amount of asynchronous collaboration. With larger operations, the teams working in RPM expand as well and are forced to rely on protocols and technology to collaborate effectively. This contrast was already evident among the two institutions of our participants, with technology mediating the majority of interactions in the larger hospital of P5 and P6. As RPM evolves and expands, this asynchronous collaboration will gain prominence. Therefore, support is needed to facilitate this collaboration, particularly when it comes to the supporting technology.  

\subsection{Foregrounding and unpacking data sensemaking in RPM}

A sensemaking activity in RPM starts with a trigger, which is usually an alarm from the monitoring app or a prompt from a colleague. This prompts the clinician to start gathering information to resolve the trigger. This activity concludes with a decision to either act, call the patient, or collaborate with a colleague. The most challenging part of the sensemaking process is aggregating all the necessary information. Information is frequently distributed across various platforms, each with its interface and data presentation method, adding layers of complexity to the retrieval process.  Interpreting this information feels meaningful once it is assembled, but locating it can be frustrating, especially when the information is in ``blocks full of free text''.

In the context of RPM, data sensemaking largely involves the hospital's EHR system, remote data-gathering platforms like Luscii \cite{noauthor_luscii_nodate}, and patient phone calls. In this ecosystem, clinicians have to bridge the gaps between different data sources through their own efforts (Figure \ref{fig:RPM collab}). They sometimes use other tools, such as digital and handwritten notes, to aid in selecting relevant data and creating summaries. While such ad-hoc strategies are helpful in the moment, they do not integrate seamlessly with the full system, and friction is added to the overall process. Our study confirms that re-contextualizing data is a critical but challenging aspect of the sensemaking process for nurses, as previously observed in related research~\cite{wright_you_2015}. This challenge is specifically relevant when the nurse needs access to qualitative information since the ``why'' and the ``how'' behind a patient's condition are often lost in long free-form text blocks. This finding supports previous studies on the use of EHRs in the hospital \cite{varpio_ehr_2015}.

The challenge extends beyond just improving the EHR systems. It encompasses the need for a holistic approach to system design that considers both the usability and accessibility of information. Such enhancements should aim to improve the presentation and coherence of data across different platforms, alongside the integration of support tools that facilitate a smoother knowledge transfer among clinicians and a smoother sensemaking process.



\subsection{Future Work}
Considering these novel insights, we see an opportunity to focus future efforts on asynchronous collaboration among clinicians. By understanding and designing for asynchronous knowledge transfer, we could mitigate the poor contextual fit or workflow disruption that often leads to implementation failures in data-driven technologies. Specifically, we see an opportunity to explore how to design tools that support the data sensemaking activities of clinicians. 

We have identified data sensemaking as an essential part of collaboration in RPM since clinicians have to encode and decode the information they wish to transfer from one another. New tools should specifically focus on facilitating this knowledge transfer between clinicians by supporting or minimizing the need to forage extensively for qualitative data.

The limitation of this exploratory study is its small sample size. With only six participants, it may not reflect the broader population of RPM clinicians. While the observations and interviews delved into their experiences, the results could lean towards the specific institutions and programs the participants belong to. Despite this limitation, our initial findings still highlight the importance of understanding asynchronous collaboration and the role of data sensemaking in it. We recommend future research to incorporate more diverse samples for a longitudinal study, and consider the experiences of a wider range of healthcare providers to validate and expand upon these findings.

%% file: main_poster/5_Conclusion.tex
\section{Conclusion}
This study identified a novel area of focus: asynchronous collaboration in Remote Patient Monitoring and the data sensemaking activities that support it. During observations, we noticed monitoring nurses are central to RPM, managing and interpreting large volumes of data. The shift to asynchronous collaboration is necessary for continuous monitoring, but existing technologies inadequately support these activities, causing friction in the daily workflow.

Current tools often increase nurses' cognitive load, due to the need to encode, decode and forage for information in the central record system. Future efforts should focus on developing technologies that enhance support for data sensemaking and facilitate smoother knowledge transfer.

In conclusion, addressing these technological and workflow challenges is vital for the effectiveness and scalability of RPM systems. Improved support for asynchronous collaboration and data sensemaking can lead to better nurse experiences and easier implementation of data-driven solutions in RPM.